\documentclass[12pt]{article}

\newcommand{\bi}{\begin{itemize}}
\newcommand{\ei}{\end{itemize}}

\usepackage{amsmath,amssymb,epsf,epsfig,amsthm,bm,graphicx,subfigure}
\usepackage[hmargin=3cm,vmargin=3cm]{geometry}

\newcommand{\boxable}{\tilde{\Box}}

\begin{document}

\author{Steven Willison\\
CENTRA, Departamento de F\'{i}sica,\\
Instituto Superior T\'{e}cnico - IST,\\
Universidade T\'{e}cnica de
Lisboa - UTL,\\ Av. Rovisco Pais 1, 1049-001 Lisboa, Portugal
}

\date{21st May 2015}

\title{Quasilinear reformulation of Lovelock gravity}

\maketitle

\begin{abstract}
 Here we give an extended review of the quasilinear reformulation of the Lovelock gravitational field equations in harmonic gauge presented in Ref. \cite{SW14}. This is important in order to establish rigorously well-posedness of the theory perturbed about certain backgrounds. The resulting system is not quasidiagonal, therefore analysis of causality is complicated in general. The conditions for the equations to be Leray hyperbolic are elucidated. The relevance to some recent results regarding the stability analysis of black holes is presented.
\end{abstract}

\section{Background}

Lovelock gravity is the most general second derivative theory in the metric consistent with the symmetry and covariant conservation of the stress tensor\cite{Lovelock}. In $d=4$ dimensions it reduces to General Relativity (with cosmological constant), but in higher dimensions there are nontrivial corrections which are higher order polynomials in the curvature tensor. It can play a role in higher dimensional theories\cite{BD} and in the context of the  AdS/CFT conjecture\cite{Edelstein}.

Here we are interested in local well-posedness of the theory. By this we mean: existence and uniqueness of solutions on some domain; continuous dependence of the solution on the data; the domain of dependence theorem holds. Essentially, for some appropriately chosen initial value surface $\Sigma$, one requires that there exists some domain $M$ so that every set of initial data (satisfying the constraints) corresponds to one and only one solution. Furthermore, two sets of initial data which agree in some region $U \subset \Sigma$ will correspond to solutions which agree in the causal domain of dependence $D^+(U)\cap M$.
The physical significance of well-posedness is that: physics is predictable; approximate solutions have physical meaning; the field has finite propagation speed. For a general nonlinear field theory, $D^+$ is determined by the principal part of the differential operator and may not coincide with that determined by the speed of light.



\section{The field equations}

The Lovelock theory field equations (in $d \geq 5$ dimensions) are of the form:
\begin{gather}
H^\nu_\mu := G^\nu_\mu + \lambda (-\frac{1}{8}\delta_{\mu \rho_1 \dots \rho_4 }^{\nu \sigma_1 \dots \sigma_4} R^{\rho_1 \rho_2}_{\quad \sigma_1 \sigma_2}R^{\rho_3 \rho_4}_{\quad \sigma_3 \sigma_4}   + \cdots) = T^\nu_\mu\, .\label{pre_Lovelock}
\end{gather}
We may alternatively write this as:
\begin{gather}
   R_{\mu\nu}+ \lambda {\cal R}_{\mu\nu}  = S_{ \mu\nu}:= T_{\mu\nu} - \frac{1}{d-2} g_{\mu\nu}\, T\, . \label{Lovelock}
\end{gather}
Above we introduced ${\cal R}_{\mu\nu}$ the ``Ricci tensor" associated with the higher order Lovelock terms. Also let $\Gamma_\mu := g_{\mu\nu}g^{\rho\sigma}
\Gamma^\nu_{\rho\sigma}$.
As a first step to obtaining a hyperbolic system of equations, one may consider the Harmonic gauge reduced equations\cite{CB_1988}.
\begin{gather}
   R_{\mu\nu}+ \lambda {\cal R}_{\mu\nu} + \partial_{(\mu} \Gamma_{\nu)} = S_{ \mu\nu}\, , \label{NL_wave}
\end{gather}
which together with the conserved initial value constraints $H_{0\mu} =T_{0\mu}$ and $\Gamma_\mu=0$ form a system equivalent to (\ref{Lovelock}).
We have a nonlinear wave equation of the general form:
\begin{gather}
  g^{\rho\sigma} \partial_\rho \partial_\sigma g_{\mu\nu} + \lambda A_{\mu\nu}(g ,\partial g,\partial \partial g) = B_{\mu\nu}(g, \partial g) \label{general_form}
\end{gather}
where $A$ is nonlinear in second derivatives. We assume for simplicity that the stress tensor is some given function. It is included in $B$.

It is clear, and well known, that the above wave equation can be put in Cauchy-Kowalevskaya form. This is because $A$ is linear in  $g_{\mu\nu, 00}$, as follows from the antisymmetrisation in (\ref{pre_Lovelock}). Existence and uniqueness for analytic data are therefore obtained in some neighbourhood of a non-characteristic initial value surface. However, it is not immediately obvious that standard well-posedness results apply to (\ref{NL_wave}) in this form. This is because, whilst they are quasilinear in time derivatives, which has many useful applications \cite{Deruelle}, they are not quasilinear in second derivatives generally. This leads to a potential loss of differentiability when attempting to apply standard fixed point theorems.

In Ref. \cite{SW14} it was pointed out that there is a method, which is quite standard in some contexts but overlooked in the literature on Lovelock theory, by which one can reformulate the equations as a quasilinear system. It is the purpose of this present article to elaborate on that method and some applications to the question of perturbations of Lovelock black holes.

\section{Obtaining a quasilinear PDE system}

By picking out the second time derivative term $M_{\mu\nu}^{\rho\sigma} g^{00} \partial_0\partial_0 g_{\rho\sigma}$, we can always put (\ref{general_form}) in the form ($\boxable := g^{\mu\nu}\partial_\mu\partial_\nu$):
\begin{gather}
 M_{\mu\nu}^{\rho\sigma} \boxable g_{\rho\sigma} = L_{\mu\nu} \, ,\label{prestart}
\end{gather}
where $M$, $L$ depend on second derivatives but not on $\partial_0\partial_0 g$.
Let us assume that $M$, viewed as a matrix of rank $n(n+1)/2$, is invertible. Then we may write:
\begin{gather}
 \boxable g_{\mu\nu} = F_{\mu\nu}(g,\partial_\mu g, \partial_i\partial_\mu g) \, .\label{start}
\end{gather}
Now we may consider the system of (\ref{start}) along with its first spatial derivatives, introducing an auxiliary field through the replacement  $\partial_i g_{\mu\nu} \to v_{\mu\nu i}$. We have
\begin{align}
 \boxable g_{\mu\nu} & =   F_{\mu\nu} (g,\partial g, \partial v) \, , \label{system1}
\\
 \boxable v_{\mu\nu i}   - \frac{\partial F_{\mu\nu} }{\partial v_{\alpha\beta j, \gamma}} \partial_\gamma\partial_j v_{\alpha\beta i}
 & = -\frac{g^{00}_{\ \ ,i}}{g^{00}} \left(F_{\mu\nu} + g^{jk}v_{\mu\nu k,j} + 2g^{j0}v_{\mu\nu j,0} \right)
 -2g^{0j}_{\ \ , i} v_{\mu\nu j,0} - g^{jk}_{\ \ ,i} v_{\mu\nu j,k}\notag
 \\ & \qquad\qquad + \frac{\partial  F_{\mu\nu} }{ \partial g_{\rho\sigma} }v_{\rho\sigma i}
 +\frac{\partial F_{\mu\nu} }{ \partial g_{\rho\sigma, \mu} } v_{\rho\sigma i,\mu}
  \, .\label{system2}
\end{align}
In (\ref{system2}) we have used (\ref{system1}) to eliminate the potentially troublesome term $g^{00}_{\ \ , i}g_{\mu\nu , 00}$.
Neglecting lower derivatives we have:
\begin{align}
 g^{\rho\sigma} \partial_\rho\partial_\sigma g_{\mu\nu} & = \cdots
\\
 g^{\rho\sigma} \partial_\rho\partial_\sigma v_{\mu\nu i}  -  \frac{\partial F_{\mu\nu}}{ \partial v_{\rho\sigma j , \mu} } \partial_\mu\partial_j v_{\rho\sigma i}
 & = \cdots
\end{align}
It is a curious fact that lower derivative terms in $L_{\mu\nu}$ contribute to the principal part of the resulting quasilinear system. In particular, since $F = M^{-1} L$ we have a term $- L_{\kappa\lambda}\frac{\partial (M^{-1})^{\kappa\lambda}_{\mu\nu}}{ \partial v_{\alpha\beta j , \gamma} } \partial_\gamma\partial_i v_{\alpha\beta j}$. We may ask whether the physical characteristics of the system really coincide with those obtained directly from the fully nonlinear system by linearising with coefficients depending on second derivatives. The linearisation of (\ref{prestart}) gives $M \boxable \delta g
=  M \delta M^{-1}\, M \boxable g + \delta L + \cdots$. So assuming the background obeys the field equations, we have $\boxable \delta g = \delta (M^{-1}L) + \cdots$. So the analysis of characteristics done in refs (\cite{Izumi,Reall}) will agree with those obtained from (\ref{system2}).

It is clear that the evolution would break down if $g^{00} =0$ or if $g$ or $M$ is not invertible. In other words, the evolution will break down if the slice $x^0 =$ const. fails to remain spacelike with respect to a well-defined metric geometry or if it becomes a characteristic surface. We shall refer to such as an acceptable initial value surface.

The initial value constraints $\phi_{\mu\nu i} := \partial_i g_{\mu\nu} - v_{\mu\nu i} = 0$, $\partial_0\phi_{\mu\nu i}= 0$ need to be imposed. We obtain the following linear homogeneous equation for $\phi$
\begin{gather}
 g^{\rho\sigma} \partial_\rho\partial_\sigma \phi_{\mu\nu i}   = -\frac{g^{00}_{\ \ ,i}}{g^{00}} \left(g^{jk}\phi_{\mu\nu k,j} + 2g^{j0}\phi_{\mu\nu j,0} \right)
 +2g^{0j}_{\ \ , i} \phi_{\mu\nu j,0} + g^{jk}_{\ \ ,i} \phi_{\mu\nu j,k}\notag
 +\frac{\partial F_{\mu\nu}}{ \partial g_{\rho\sigma, 0} } \phi_{\rho\sigma i,0}    \, .
\end{gather}
The above expresses the conservation of the constraints, and will be hyperbolic if (\ref{system1}) and (\ref{system2}) are. Therefore, given an acceptable initial value surface, with data satisfying all the constraints, solutions of (\ref{system1}) and (\ref{system2}) are equivalent to solutions of Lovelock equations (in harmonic gauge) on the domain of dependence.

Since (\ref{system1}) and (\ref{system2}) constitute a quasilinear PDE system, it remains to check whether they are indeed of some hyperbolic type, such that the domain of dependence is non-trivial.
The system is not quasidiagonal, so some work is required. But, since the offending term $-  \frac{\partial F_{\mu\nu}}{ \partial v_{\rho\sigma j , \mu} } \partial_\mu\partial_j v_{\rho\sigma i}$ is of order $\lambda$, for small perturbations about flat spacetime, we may expect that the system is (Leray) hyperbolic and the characteristics approximately light-like as expected.

In the case that the curvature is large, it is not so clear. Looking at (\ref{system2}), we are lead to consider the following characteristic matrix:
\begin{gather}
p^A_B(X,X) = \delta^A_B g^{\mu\nu}X_\mu X_\nu + \frac{\partial F^A}{\partial v_{Bj,\rho}} X_j X_\rho \, ,
\end{gather}
which is a quadratic form on the co-tangent space at point $x \in M$. The characteristic determinant $P^*_x = \det p^A_B$ is a homogeneous polynomial of degree $m= d(d+1)$. The equation $P^*_x(X) = 0$ defines a cone in the co-tangent space. A homogeneous polynomial of degree $m$ is hyperbolic if the cone has the following property: there exists a point $P$ in $T^*_x$ such that any straight line passing through $P$ which does not pass through the vertex, cuts the cone exactly $m$ times. The differential operator $p^A_B(\partial,\partial)$, is hyperbolic at $x$ if $P^*_x$ is hyperbolic or if it is a product of hyperbolic polynomials.

A differential operator hyperbolic at $x$ defines a convex causal cone in the tangent space $T_x$. This can be used to define the notion of causal paths. The operator is said to be \emph{globally hyperbolic} on a manifold $M$ if the set of causal paths is a compact subspace of the space of all paths, in the appropriate topology\cite{CBbook}. If the operators $p^A_B(\partial,\partial)$ and $g^{\mu\nu}\partial_\mu\partial_\nu$ are globally hyperbolic on $M$, then the Lovelock equations in harmonic gauge are of Leray type. Existence, uniqueness and domain of dependence properties then follow by known results. Continuous
dependence on the data is known for some cases\cite{CBbook}.

Since the coefficients depend in a complicated way on the field itself, verifying hyperbolicity is in general a difficult business.
There are two ways to attempt to further simplify. One is to attempt by change of variables to quasidiagonalise the equations, or to make it quasidiagonal by blocks i.e. $p^A_B(\partial,\partial) = f_A \delta^A_B \boxable $. Another is to reduce to a first order system.

\section{Characteristic surfaces and black holes in Lovelock theory}

The spherically symmetric black hole solution of quadratic Lovelock gravity (Einstein-Gauss-Bonnet or EGB gravity) is that of Boulware and Deser\cite{BD}. The solution has two branches, only one of which is believed to be physically meaningful, due to the ghost instability of the bad branch. The relevant facts for our purposes are that the causal diagram looks the same as for Schwarzchild-Tangherlini, the event horizon and (assuming zero cosmological constant) conformal infinity are both null.

It was shown by Izumi \cite{Izumi} that the event horizon is also a characteristic surface of the EGB equations. Furthermore, it is an immediate consequence of asymptotic flatness that conformal infinity is characteristic in an appropriate limiting sense. The same statements hold for the analogous Lovelock black holes\cite{Reall}. Therefore one might expect that these form the boundary of a causal domain of dependence, as they do in General Relativity. This is indeed a reasonable expectation assuming that the local causal structure varies in a continuous fashion in the interior.

Reall, Tanahashi and Way\cite{Reall} considered the Lovelock equations for perturbations about the black hole background, transforming them to a quasidiagonal-by-blocks form by introducing new variables corresponding to scalar, vector and tensor perturbations (c.f. \cite{Dotti}). Heuristically:
\begin{align*}
 G_S^{\mu\nu} \partial_\mu\partial_\nu \psi_S & = \cdots \, ,
\\
 G_V^{\mu\nu} \partial_\mu\partial_\nu \psi_V & = \cdots \, ,
\\
 G_T^{\mu\nu} \partial_\mu\partial_\nu \psi_T & = \cdots \, .
\end{align*}
Then hyperbolicity amounts to the effective metrics $G_S$, $G_V$ and $G_T$ having Lorentzian signature. It was found that for some small black holes, there was a region near to the horizon where the Lorentzian character broke down for one of the degrees of freedom. Therefore for these black holes, the exterior region is not globally hyperbolic and the perturbation problem is not well-posed. They argued that for large black holes the effective metrics remained Lorentzian. Therefore, we would expect the perturbation problem to be well-posed in these cases. Our quasilinear reformulation of the theory can be regarded as an additional justification for such an expectation.

\textbf{Acknowledgements:} I thank H. Reall for a helpful comunnication. This work was supported in its early stages by the
Funda\c{c}\~{a}o para a Ci\^{e}ncia e a Tecnologia of Portugal and
the Marie Curie Action COFUND of the European Union Seventh
Framework Programme (grant agreement PCOFUND-GA-2009-246542).


\begin{thebibliography}{99}
\bibitem{Lovelock}
  D.~Lovelock,
  J.\ Math.\ Phys.\  {\bf 12}, 498 (1971).

\bibitem{BD}
  D.~G.~Boulware and S.~Deser,
  Phys.\ Rev.\ Lett.\  {\bf 55}, 2656 (1985).

\bibitem{Edelstein}
  X.~O.~Camanho and J.~D.~Edelstein,
  JHEP {\bf 1006}, 099 (2010)
  [arXiv:0912.1944 [hep-th]];
  X.~O.~Camanho, J.~D.~Edelstein and M.~F.~Paulos,
  JHEP {\bf 1105}, 127 (2011)
  [arXiv:1010.1682 [hep-th]].


  \bibitem{CB_1988}
  Y.~Choquet-Bruhat,
  ``Gravitation with a Gauss Bonnet term,''
  Australian National University Publications, R. Bartnick ed. , 53 (1988),
15
Republished in \cite{CBbook};``The Cauchy Problem for Stringy Gravity,''
  J.\ Math.\ Phys.\  {\bf 29}, 1891 (1988).

\bibitem{CBbook}
Y. Choquet-Bruhat, General Relativity and the Einstein Equations, Oxford,
2009, pp. 689-708.

\bibitem{Deruelle}
  N.~Deruelle and J.~Madore,
  gr-qc/0305004.

\bibitem{SW14}
  S.~Willison,
  Class.\ Quant.\ Grav.\  {\bf 32}, no. 2, 022001 (2015)
  [arXiv:1409.6656 [gr-qc]].

\bibitem{Izumi}
  K.~Izumi,
  Phys.\ Rev.\ D {\bf 90}, no. 4, 044037 (2014)
  [arXiv:1406.0677 [gr-qc]].
\bibitem{Reall}
  H.~Reall, N.~Tanahashi and B.~Way,
  Class.\ Quant.\ Grav.\  {\bf 31}, 205005 (2014)
  [arXiv:1406.3379 [hep-th]].

\bibitem{Dotti}
  R.~J.~Gleiser and G.~Dotti,
  Phys.\ Rev.\ D {\bf 72}, 124002 (2005)
  [gr-qc/0510069].


\end{thebibliography}
\end{document}